\documentclass[12pt]{article}
\usepackage{amssymb}
\usepackage{epsfig}

\catcode`\@=11
\def\numberbysection{\@addtoreset{equation}{section}
        \def\theequation{\thesection.\arabic{equation}}}
\def\beq{\begin{equation}}
\def\eeq{\end{equation}}
\numberbysection
\begin{document}
\begin{titlepage}
\begin{center}
\hfill DFF  1/7/99 \\
\vskip 1.in
{\Large \bf Perturbative Analysis of the Two-Body Problem in
($2+1$)-$AdS$ Gravity}
\vskip 0.5in
P. Valtancoli
\\[.2in]
{\em Dipartimento di Fisica dell' Universita', Firenze \\
and INFN, Sezione di Firenze (Italy)}
\end{center}
\vskip .5in
\begin{abstract}
We derive a perturbative scheme to treat the interaction between point
sources and $AdS$-gravity. The interacting problem is equivalent to
the search of a polydromic mapping  $X^A=X^A(x^\mu)$ ,
endowed with $O(2,2)$ monodromies, between the physical coordinate
system and a Minkowskian 4-dimensional coordinate system, which is
however constrained to live on a hypersurface. The physical motion
of point sources is therefore mapped to a geodesic motion on this
hypersurface. We impose an instantaneous gauge which induces a set of
equations defining such a polydromic mapping. Their consistency leads
naturally to the Einstein equations in the same gauge. We explore the
restriction of the monodromy group to $O(2,1)$, and we obtain the solution
of the fields perturbatively in the cosmological constant.
\end{abstract}
\medskip
\end{titlepage}
\pagenumbering{arabic}
\section{Introduction}

The scattering of particles in gravity is one of the typical problems
of physics, from the times relativity theory has been discovered
\cite{a1}-\cite{a2}-\cite{a3}-\cite{a4}.
Our scheme of solution \cite{a5}-\cite{a6}-\cite{a7}-\cite{a8}
 is based on introducing the constants of motion
of the problem by allowing for a polydromic mapping $X^A = X^A(x^\mu)$
of the flat metric,
whose polydromy automatically builds up the delta singularities of the
particles' energy-momentum tensors.

Thanks to this mapping the physical motion can be trivialized to an
elementary motion, which in the case of pure gravity reduces to a free
motion on a plane, while, as it has been shown in \cite{a9}, in the case of $AdS$
gravity reduces to a geodesic motion of a hyper-surface immersed in a
flat 4-dimensional Minkowskian space-time with signature ( + + - - ). In
the last scheme we can unify not only particle-like singularities,
moving as time-like geodesics, but also $BTZ$ black-hole singularities
\cite{a10}-\cite{a11}-\cite{a12}-\cite{a13}, moving as space-like geodesics.

In this article we restrict such a mapping problem in a particular
gauge, the instantaneous gauge defined by the equations $g_{zz} = 0 $
and  $K=0$, which has allowed to compute, in the case of pure
$(2+1)$-gravity \cite{a6}, the fields in terms of analytic
and anti-analytic functions. Our aim is to show how this scheme can be
extended to the case of $(2+1)$ $AdS$-gravity \cite{a14}-\cite{a15},
although we are not able
to solve completely for the fields. Our method at least gives a framework
in which it is possible to obtain perturbatively an unique answer.

Such a polydromic mapping is constrained by the gauge choice to
satisfy a complete set of equations governing this immersion in the flat
4-dimensional space-time. This system of equations is related to the
gauge choice and is valid only for those regions of space-time consistent
with the conformal gauge. An obvious counterexample is the case of the
region of the black hole internal to the horizon, whose metric is too complex
to satisfy such a choice.

The first-order formalism is simpler than the second-order formalism,
as it allows to avoid distributions, related to the singularities and
it instead introduces polydromies which take into account globally the
presence of the particles.

In this sense sources are viewed as singularities of the fields
once that the monodromy conditions are imposed around them. The
particles' scattering is then reduced to a composition of monodromies,
that, in the case of $O(2,2)$ monodromy group, produces two invariant
masses \cite{a9},
determining the solution at great distances. In the particular
case of zero angular momentum, these two invariant masses coincide
between them and the monodromy group is restricted to $O(2,1)$
( see also \cite{a16} ). In
such a case it is useful to introduce a parameterization that reduces
the cut to a M\"{o}bius transformation. However we are not able, as in the
gravity case, to find a parameterization of the fields in terms of
analytic functions, and here the problem seems quite involved, being
the equations for the $X^A$ coordinates non-analytic.

It is however possible to solve for the cuts with non-analytic
functions, since the non-analytic part can be perturbatively represented as a
finite number of terms, obtained as products of analytic and
anti-analytic functions. It remains an arbitrariness of analytic
functions in the naive integration that is removed  imposing that
the whole solution satisfies the monodromies. We obtain therefore,
order by order in perturbation theory, a field satisfying all the
requirements and depending from the constants of motion, which
describe the particles' motion. We finally discuss the limits of such
research to the study of black holes scattering.

\section{Instantaneous gauge in the second order formalism}

The splitting ADM of space-time in terms of space-like surfaces
can be derived from the Einstein-Hilbert action with negative
cosmological constant ( with modulus $\Lambda$ )
 rewriting the scalar curvature $R^{(3)}$ in its
spatial part $R^{(2)}$, intrinsic to the surfaces $\Sigma (t)$,
and an extrinsic part, coming from the embedding, as follows ( $ 8 \pi
G = 1 $ )
\beq S = - \frac{1}{2} \int \ \sqrt{| g |} R^{(3)} + \Lambda \int \sqrt{| g |}
= - \frac{1}{2} \int \ \sqrt{| g |}
[ R^{(2)} + {( Tr K )}^2 - Tr ( K^2 ) ]
+ \Lambda \int \sqrt{| g |} . \label{a1}\eeq
where the equivalence is true apart from a surface term. Here we have
introduced the extrinsic curvature tensor $K_{ij}$, o second
fundamental form of the surface $\Sigma (t)$, given by
\beq  K_{ij} = \frac{1}{2} \sqrt{ \frac{|g_{ij}|}{|g|} } \left( \nabla^{(2)}_i
g_{0j} + \nabla^{(2)}_j g_{0i} - \partial_0 g_{ij}  \right) ,  \label{a2}\eeq
with the notation that the spatial indices are raised or lowered with
the spatial part of the metric $g_{ij}$ and its inverse.

Let us choose as parameterization for the metric, once that it is
constrained to be in conformal gauge $g_{zz} = 0$:
\beq g_{00} = \alpha^2 - e^{2\phi} \beta \overline{\beta} , \ \ \ \
g_{0z} = \frac{1}{2} \overline{\beta} e^{2\phi}, \ \ \ \
g_{z \overline{z}} = - \frac{1}{2} e^{2\phi} . \label{a3}\eeq
Let us compute the trace of the extrinsic curvature tensor and pose it
equal to zero, in such a way that all the dependences on time
derivatives are avoided in the action :
\beq  K ( z, \overline{z}, t ) = K_{z \overline{z}} =
\frac{1}{2 \alpha} ( \partial_z g_{0\overline{z}} +
\partial_{\overline{z}} g_{0z} - \partial_0 g_{ z \overline{z}} ) = 0 .
\label{a4}\eeq

The other part of the tensor $K_{ij}$ can be shown to be a meromorphic
analytic function, as we shall see in a moment from the equations of motion
\beq K_{zz} = N(z) = \frac{e^{2\phi}}{2\alpha} \partial_z
\overline{\beta} . \label{a5}\eeq
The condition of instantaneous gauge ( see also \cite{a6} )
is therefore defined by
\beq g_{zz} = K = 0 . \label{a6}\eeq

Combining the conditions above, we obtain a new action without
temporal derivatives, showing that the propagation of the fields
$\alpha$, $\beta$, $\phi$ can be made instantaneous. Adding the
matter coupling (i.e. a set of $N$ particles ), the equations of
$AdS$-gravity have as sources both the cosmological constant that can
be thought as the mean effect of an uniform distribution of matter and
the energy-momentum tensor of the external particles:
\begin{eqnarray}
\nabla^2 \phi + \frac{e^{2\phi}}{\alpha^2} \partial_z \overline{\beta}
\partial_{\overline{z}} \beta & = & \nabla^2 \phi + 4 N \overline{N} e^{-2\phi}
= \Lambda e^{2\phi} - |g| e^{-2\phi} T^{00} \nonumber \\
\partial_{\overline{z}} \left( \frac{e^{2\phi}}{2 \alpha} \partial_z
\overline{\beta} \right) & = & \partial_{\overline{z}} N(z) =
- {( 4 \alpha )}^{-1} |g| ( T^{0z} - \beta T^{00} ) \nonumber \\
\nabla^2 \alpha - 2 \frac{e^{2\phi}}{\alpha^2} \partial_z \overline{\beta}
\partial_{\overline{z}} \beta & = & \nabla^2 \alpha -  8 N
\overline{N} e^{-2\phi} \alpha = 2 \Lambda \alpha e^{2\phi} +
\alpha^{-1} |g| ( T^{z\overline{z}} - \beta T^{0\overline{z}} -
\overline{\beta} T^{0z} + \beta \overline{\beta} T^{00} ) , \nonumber \\
& &  \label{a7}\end{eqnarray}
where $\nabla^2 \equiv 4\partial_z \partial_{\overline{z}}$ denotes
the Laplacian, and the energy-momentum tensor for the particles is
given by
\beq T^{\mu\nu} = \frac{1}{\sqrt{|g|}} \sum_{(i)} m_i \left(
\frac{dt}{ds_i} \right) \dot{\xi}^\mu_i \dot{\xi}^\nu_i
\delta^2 ( \underline{x} - \underline{\xi}_i (t_i) , \ \ \ \ \ \
(i=1,2,...,N). \label{a8}\eeq

Avoiding time derivatives allows to treat the particle
motion as an external independent datum, from which the metric has a parametric
dependence, while only by imposing both the monodromies and the boundary
conditions one obtains constraints on the geodesic motion of the particles.

With the introduction of the conformal gauge $g_{zz}=0$
the Einstein equations for the corresponding components of the Ricci
tensor $R_{\mu\nu}$ are missing in eq. (2.7), and should be added as a
constraint,i.e.
\beq R_{zz} = T_{zz} . \label{a9}\eeq
These two constraints, together with the gauge condition $K= 0$, are
indeed equivalent to the covariant conservation of the
energy-momentum tensor, which in turn implies the geodesic equations,
as shown in ref. \cite{a6}.

\section{Comparison with the first-order formalism}

The introduction of the cosmological constant makes useless searching
for a simplification of the equations of motion in the dreibein
formalism \cite{a14},
while it is fruitful to start from the typical construction
of space-times with constant curvature \cite{a10}-\cite{a11},
that are obtained from a
4-dimensional flat metric with a constraint between the coordinates:
\beq ds^2 = dX^A dX^B \eta_{AB} \ \ \ \ \ \ X^A X_A =
\frac{1}{\Lambda} . \label{b1}\eeq
This formalism can also be considered a first-order formalism,
defining a new dreibein $E^A_\mu = \partial_\mu X^A$, in
which the Lorentz index  depends on four coordinates.

From the constraint (\ref{b1}) we can easily deduce the following properties:

\beq X^A \partial_z X_A = 0 \ \ \ \ X^A \partial_0 X_A = 0 . \label{b2}\eeq

If we suddenly impose the conformal gauge, i.e.
\beq g_{zz} = \partial_z X^A \cdot \partial_z X_A = 0 \ \ \ \ \ \ \ \
g_{z\overline{z}} = \partial_z X^A \cdot \partial_{\overline{z}} X_A =
-\frac{1}{2} e^{2\phi} , \label{b3}\eeq
many other properties of the $X^A$-mapping are shown to be valid.

An intrinsic local frame of the four-dimensional space-time is given by
these 4 four-vectors $ X^A,
\partial_z X^A , \partial_{\overline z} X^A , $ and $V^A$, defined as
\beq V^A = 2i \sqrt{\Lambda} e^{-2\phi}  \epsilon^{A B C D} X_B \partial_z X_C
\partial_{\overline z} X_D \label{b4}\eeq
that is orthogonal to the first three ones, and is defined to have
norm equal to unity. Instead $\partial_z X^A$ and
$\partial_{\overline{z}} X^A$ are two null vectors that have a
non-vanishing scalar product between them. We can easily check the
following property:
\beq \partial_0 X^A = - \beta \partial_z X^A - \overline{\beta}
\partial_{\overline{z}} X^A + \alpha V^A . \label{b5}\eeq
Therefore once that the mapping problem $X^A$ is solved at fixed time,
we can also determine the field  $\alpha$ and $\beta$, while the
knowledge of the field $\phi$ doesn't require to compute any time
derivative on the mapping.

In the second order formalism we have chosen an instantaneous gauge,
which is the same that we have use to solve the $N$-body problem in
pure ($2+1$)-gravity, and it is defined by the conditions in eq. (\ref{a6}).
These gauge conditions combined together imply that :
\beq  \partial_z \partial_{\overline{z}} X^A = \frac{\Lambda}{2}
e^{2\phi} X^A . \label{b6}\eeq
In particular, the $K=0$ condition implies that additional terms
proportional to $V^A$ are absent in the r.h.s. of eq. (\ref{b6}).

Let us compute $\partial_z^2 X^A$. We can develop it in the basis of
four-vectors
\beq \partial^2_z X^A =  c_0 X^A + c_1 \partial_z X^A +
c_2 \partial_{\overline{z}} X^A + c_3  V^A .
\label{b7}\eeq
It is easy to see that $c_0 = 0$ and $c_2 = 0$ are consequences of the
eqs. (\ref{b2}) and (\ref{b3}).
The scalar product with  $\partial_{\overline{z}} X^A$
implies that $c_1 = 2 \partial_z \phi$, while $c_3$ can be constrained
since it satisfies the analyticity condition:
\beq \partial_{\overline{z}} c_3 = 0 , \label{b8}\eeq
therefore $c_3 = c_3 (z)$ is a monodromic analytic function. The only
candidate for such a function is the meromorphic function $N(z)$,
previously identified as the $K_{zz}$ component of the extrinsic
curvature tensor; in fact, starting from the definition (\ref{a5}) of $N(z)$
we can show that  $c_3 = N(z)$. Therefore we find the equation:
\beq \partial^2_z X^A = 2 \partial_z \phi \partial_z X^A + N(z) V^A .
\label{b9}\eeq
Now let us compute $\partial_z V^A$. From the definition of $V^A$
we can suddenly notice that $\partial_z V^A$ is orthogonal both to $V^A$ and
$X^A$. We can therefore pose
\beq \partial_z V^A = a_1 \partial_z X^A + a_2 \partial_{\overline{z}}
X^A .\label{b10}\eeq
Introducing eq. (\ref{b9}) in the computation of (\ref{b10}) we can
directly find $\partial_z V^A$, and obtain that:
\beq \partial_z V^A = 2 e^{-2\phi} N(z) \partial_{\overline{z}} X^A .
\label{b11}\eeq
In summary, since we work in an instantaneous gauge, the equations at
fixed time of the immersion $X^A= X^A(x^\mu)$ are given by:
\begin{eqnarray}
\partial_z \partial_{\overline{z}} X^A & = & \frac{\Lambda}{2} e^{2\phi} \
X^A \nonumber \\
\partial^2_z X^A & = & 2 \partial_z \phi \partial_z X^A + N(z) V^A
\nonumber \\
\partial_z V^A & = & 2 e^{-2\phi} N(z) \partial_{\overline{z}} X^A .
\label{b12}\end{eqnarray}
The consistency of these equations leads to the fields equations
(\ref{a7}) in the second-order formalism.

Let us also notice the following property,i.e., that, being $X^A$ and
$V^A$ two vectors of constant norm, their dynamics can be described by a
sigma model $O(2,2)$, as a consequence of the equations (\ref{b12})
\begin{eqnarray}
\partial_z \partial_{\overline{z}} X^A & = & - \Lambda
( \partial_z X^B \partial_{\overline{z}} X_B )) X^A \nonumber \\
\partial_z \partial_{\overline{z}} V^A & = & - ( \partial_z V^B
\partial_{\overline{z}} V_B )) V^A .
\label{b13}\end{eqnarray}
Both eqs. (\ref{b12}) and (\ref{b13}) are automatically covariant with
respect to the $O(2,2)$ cuts, that can be decomposed as products of
$SU(1,1)$ cuts, as shown in ref. \cite{a9}. In general, the one-particle
cut can be identified as ( $ X^A = ( X^t , \overline{X}^t , X^z ,
\overline{X}^z ) $ )
\beq \left( \begin{array}{cc} X^t & X^z \\ \overline{X}^z & \overline{X}^t
\end{array} \right) \rightarrow
\left( \begin{array}{cc} A_1 & B_1 \\ \overline{B}_1 & \overline{A}_1
\end{array} \right)
\left( \begin{array}{cc} X^t & X^z \\ \overline{X}^z & \overline{X}^t
\end{array} \right)
\left( \begin{array}{cc} A_2 & B_2 \\ \overline{B}_2 & \overline{A}_2
\end{array} \right) \label{b14}\eeq
where the entries $A_i$ and $B_i$ of the monodromy matrices are given by:
\begin{eqnarray}
A_1 \ & = & \ cos \pi \mu  - i ch ( \lambda_1 - \lambda_2 ) sin \pi \mu
 \nonumber \\
B_1 \ & = & \ - i e^{- i( \alpha + \beta )} sh ( \lambda_1 - \lambda_2 )
sin \pi \mu \nonumber \\
A_2 \ & = & \ cos \pi \mu  + i ch ( \lambda_1 + \lambda_2 ) sin \pi \mu
 \nonumber \\
B_2 \ & = & \ - i e^{ i( \alpha - \beta )} sh ( \lambda_1 + \lambda_2 )
sin \pi \mu .
\label{b15} \end{eqnarray}

In general the solution of the two-body problem can be obtained
in global terms. The result of the composition of two
monodromies, in the case of particles, is of course of the type:

\beq \left( \begin{array}{cc} {X'}^t & {X'}^z \\ \overline{X'}^z &
\overline{X'}^t \end{array} \right) =  M_L
\left( \begin{array}{cc} X^t & X^z \\ \overline{X}^z &
\overline{X}^t \end{array} \right) M_R ,
\label{b16} \eeq
where to the $M_L, M_R$ matrices it is possible to associate the
corresponding invariant masses \cite{a4}:

\begin{eqnarray}
cos ( \pi {\cal M}_L ) & = & cos (\pi \mu_1 ) cos ( \pi \mu_2 ) - \frac{P_1^L
\cdot P_2^L}{m_1 m_2} sin ( \pi \mu_1 ) sin ( \pi \mu_2 ) \nonumber \\
cos ( \pi {\cal M}_R ) & = & cos (\pi \mu_1 ) cos ( \pi \mu_2 ) - \frac{P_1^R
\cdot P_2^R}{m_1 m_2} sin ( \pi \mu_1 ) sin ( \pi \mu_2 )
\label{b17}\end{eqnarray}
and we have defined the following vectors, constants of motion:

\begin{eqnarray}
P_1^L & = & m_1 \gamma_1^L ( 1, v^L_1 ) \ \ \ \ \
\gamma_1^L = ch ( \lambda_1 - \lambda_2 ) \ \ \
\gamma_1^L v_1^L = e^{-i (\alpha+\beta)} sh ( \lambda_1 - \lambda_2 )
\nonumber \\
P_1^R & = & m_1 \gamma_1^R ( 1, v^R_1 ) \ \ \ \ \
\gamma_1^R = ch ( \lambda_1 + \lambda_2 ) \ \ \
\gamma_1^R v_1^R = e^{-i (\alpha-\beta)} sh ( \lambda_1 + \lambda_2 ) .
\label{b18} \end{eqnarray}

For generic values of the constants of motions, the
left invariant mass ${\cal M}_L$ will be different from the
right invariant mass ${\cal M}_R$ and therefore the composed system
has spin, other than invariant mass.

In the simplified case in which $A_1 = \overline{A}_2 = A$,  $B_1 = B_2 =
B$, that correspond to null angular momentum, it is useful to
parameterize the $X^A$ transformation in the following way, that makes
explicit the reduction of monodromies from the generic case of
$O(2,2)$ to the particular case of $O(2,1)$  ( $X^A = ( X^0, X^1, X^z
, \overline{X}^z ) $ )
\beq
X^A = \frac{1}{\sqrt{\Lambda}} \left( \frac{1+ Z \overline{Z}}{1-
Z \overline{Z}} cos T , sin T, \frac{ 2 Z }{1-
Z \overline{Z}} cos T ,  \frac{ 2 \overline{Z} }{1-
Z \overline{Z}} cos T \right) , \label{b19}\eeq
in which the only polydromic variable is  $Z$, transforming as $O(2,1)$
around the particles
\beq Z \rightarrow \frac{A Z + B }{\overline{B} Z + \overline{A}} .
\label{b20}\eeq
Exploring the consequences of eqs. (\ref{b12}) we find that there are two
eqs. with a Laplacian for the $T$ and $Z$ variables:
\begin{eqnarray}
\partial_z \partial_{\overline{z}} T & = & 2 sin T cos T \left(
\frac{ \partial_z Z \partial_{\overline{z}} \overline{Z} +
 \partial_z \overline{Z} \partial_{\overline{z}} Z }{ {( 1 - Z
\overline{Z} )}^2 } \right) \nonumber \\
\partial_z \partial_{\overline{z}} Z & + & \frac{ 2 \overline{Z}
\partial_z Z \partial_{\overline{z}} Z }{ ( 1 - Z \overline{Z} )}
 = \frac{2 sin T \sqrt{ \partial_z Z \partial_{\overline{z}} Z }}{
( 1 - Z \overline{Z} )} \left( \sqrt{ \partial_z \overline{Z}
\partial_{\overline{z}} Z } + \sqrt{ \partial_z Z
\partial_{\overline{z}} \overline{Z} } \right) .
\label{b21}\end{eqnarray}
The first one is integrated by the following first integral,
\beq \partial_z T = \frac{2 \sqrt{\partial_z Z \partial_z
\overline{Z}} }{ 1 - Z \overline{Z} } cos T , \label{b22}\eeq
that is also obtained developing the condition of null norm
 $\partial_Z X^A \cdot \partial_Z X_A = 0 $.

The second one is integrated solving the eqs. (\ref{b12})
in terms of $N(z)$:
\beq N(z) = \frac{1}{\sqrt{\Lambda}} \left( \frac{ \sqrt{\partial_z Z
\partial_{\overline{z}} \overline{Z} } - \sqrt{\partial_z \overline{Z}
\partial_{\overline{z}} Z } }{ \sqrt{\partial_z Z
\partial_{\overline{z}} \overline{Z} } + \sqrt{\partial_z \overline{Z}
\partial_{\overline{z}} Z } } \right)
\frac{e^{2\phi}}{cos T} \partial_z ( e^{-2\phi}
\partial_z sin T ) . \label{b23}\eeq

Using these first integrals we can express the conformal factor
$e^{2\phi}$ in such a way that makes explicit its invariance under $O(2,1)$:

\beq e^{2\phi} = \frac{4}{\Lambda} { \left( \frac{ \sqrt{ \partial_z Z
\partial_{\overline{z}} \overline{Z} } -
\sqrt{ \partial_z \overline{Z}
\partial_{\overline{z}} Z } }{
( 1- Z \overline{Z} ) } \right) }^2 cos^2 T . \label{b24}\eeq
This method appears to be more involved than the second-order
formalism, however the latter one would require to analyze in detail
all the various contributions, of distribution type, while introducing
the cuts allows to avoid to write down mathematically ill-defined equations.

\section{Perturbative expansion in $\Lambda$ for two-body scattering  }

Let us firstly recall the single body metric, which is defined in the
radial gauge as
\beq ds^2 = ({(1-\mu)}^2 + \Lambda r^2 ) dt^2 -
\frac{ dr^2}{{(1-\mu)}^2 + \Lambda r^2} - r^2 d\theta^2 \label{c1}\eeq
This metric can be expressed as a polydromic mapping
\begin{eqnarray}
X^t & = & X^0 + i X^1 = \frac{1}{\sqrt{\Lambda}} \sqrt{ 1 + \frac{
\Lambda r^2}{{(1-\mu)}^2 } } e^{i\sqrt{\Lambda} ( 1-\mu) t} \nonumber \\
X^z & = & X^2+i X^3 = \frac{r e^{i(1-\mu) \theta}}{(1-\mu)} .
\label{c2}\end{eqnarray}
Introducing the spatial conformal gauge $g_{zz}=0$, which is
obtained with a radial coordinate transformation
\beq r \rightarrow \frac{(1-\mu) r^{(1-\mu)} }{1-\frac{\Lambda}{4}
r^{2(1-\mu)} } , \label{c3}\eeq
the $X^A$-mapping becomes
\begin{eqnarray}
X^t & = & \frac{1}{\sqrt{\Lambda}} \frac{ 1 + \frac{\Lambda}{4}
r^{2(1-\mu)} }{ 1 - \frac{\Lambda}{4} r^{2(1-\mu)} }
e^{i\sqrt{\Lambda} ( 1-\mu) t} \nonumber \\
X^z & = & \frac{z^{1-\mu} }{  1 - \frac{\Lambda}{4} r^{2(1-\mu)} }.
\label{c4}\end{eqnarray}
In conformal gauge it appears a physical limit on the radial
coordinate, implying that the solution is defined on a disk instead on
the whole $z$-plane, and it is given by  $ r^{1-\mu} \leq
\frac{2}{\sqrt{\Lambda}} $.

Before starting the study of particles' scattering, let us analyze in
detail the one-body problem. Firstly, we can say that $N(z)=0$, implying
by consistency with eq. (\ref{b23}) that
\beq e^{-2\phi} \partial_z sin T = c(t) \overline{z} . \label{c5}\eeq
From the one-body problem solution (\ref{c4}), the following expressions
for $T$ and $Z$ have been obtained in \cite{a9}
\begin{eqnarray}
Z & = & \sqrt{1-a^2} \frac{\sqrt{\Lambda}}{4}  z^{1-\mu} \left[
1 + \frac{4 r^{-2(1-\mu)}}{\Lambda} \left(
1- \sqrt{ 1 - \frac{\Lambda}{2} \left( \frac{ 1+a^2 }{ 1-a^2 } \right)
 r^{2(1-\mu)} + \frac{\Lambda^2}{16}  r^{4(1-\mu)} } \right) \right] \sim
\nonumber \\
& \sim & \frac{\sqrt{\Lambda}}{2} \frac{z^{1-\mu}}{\sqrt{1-a^2}} \left[ 1 +
\frac{\Lambda}{4} \frac{a^2}{1-a^2} r^{2(1-\mu)} + O(\Lambda^3 )
\right] \nonumber \\
sin T & =& a \frac{1 + \frac{\Lambda}{4} r^{2(1-\mu)} }{
1 - \frac{\Lambda}{4} r^{2(1-\mu)} } \ \ \ \ \ \ \ \ a = sin (
\sqrt{\Lambda} (1-\mu) t ) .
\label{c6}\end{eqnarray}
Therefore we can fix the time-dependent constant of eq. (\ref{c5}) as
\beq c(t) = \frac{\Lambda}{2(1-\mu)} a(t) . \label{c7}\eeq

This solution is not valid with the choice of the parameterization (\ref{b19})
in the whole disk, but only in a subset defined by the condition $|Z|
\leq 1$, that in the case of a single body is also a disk in the
physical coordinates defined by
\begin{eqnarray}
 r^{2(1-\mu)} & \leq & \frac{4}{\Lambda} \frac{1-a(t)}{1+a(t)}
\ \ \ {\rm \ for \ \ } a(t) > 0 \nonumber \\
& \leq & \frac{4}{\Lambda} \frac{1+a(t)}{1-a(t)}
\ \ \ {\rm \ for \ \ } a(t) < 0 .
\label{c8}\end{eqnarray}
At the limiting value, the $X^A$ mapping reduces to  $\Theta = (1-\mu)\theta$.
The continuation of $Z$-field outside the physical region meets
another region, disconnected from the physical one, in which the
condition $|Z| \leq 1$ is valid. In this specular universe there is also an
image of the source. In the two-body case this image will remain an
isolated singularity, whose cut will be identified with the
composition of the two-body cut.

Let us notice that this limit on the coordinates is not perturbative
in $\Lambda$ and that at each order of the perturbative expansion this
limit can be considered $\infty$. Only after re-summing the
perturbative series, we can understand what is precisely the domain
of the exact solution, from which it must be continued with another
choice of the parameterization, instead of eq. (\ref{b19}).

With respect to the geodesic motion of a test particle situated in
$\xi(t)$, under the influence of a mass source situated in the origin of
the coordinates, the knowledge of this patch is enough to give the
complete solution, if the $Z$-value corresponding to $\xi(t)$ has
modulus less than unity:
\begin{eqnarray}
 Z(\xi) & = & th ( \sqrt{\Lambda} \lambda ) \nonumber \\
\sqrt{\Lambda} r^{1-\mu} & = & \frac{1}{\sqrt{1-a^2(t)}} \left[
\left( Z(\xi) + \frac{1}{Z(\xi)} \right) -
\sqrt{ {\left( Z(\xi) + \frac{1}{Z(\xi)} \right)}^2 - 4 ( 1 - a^2 (t)
) } \right] .
\label{c9}\end{eqnarray}
The resulting motion is limited on a line, similar to the
harmonic motion, and the test particle can meet the source after a
finite time $t_0$.

To introduce the two-body problem, it is useful to give the explicit
development of eq. (\ref{c6}) at the first orders:
\begin{eqnarray}
T & = & \sqrt{\Lambda} ( T^{(0)} + \Lambda  T^{(1)} + O(\Lambda^2) )
\nonumber \\
& = & \sqrt{\Lambda} (1-\mu) t \left( 1 + \frac{\Lambda}{2}
 r^{2(1-\mu)}   + O({\Lambda}^2) \right) \nonumber \\
Z & = & \frac{\sqrt{\Lambda}}{2} ( Z^{(0)} + \Lambda  Z^{(1)} +
\Lambda^2  Z^{(2)} + O(\Lambda^3) ) \nonumber \\
& = & \frac{\sqrt{\Lambda}}{2} z^{1-\mu} \left( 1 +
\frac{\Lambda}{2} {(1-\mu)}^2 t^2 +
\frac{5 \Lambda^2}{24} {(1-\mu)}^4 t^4 +
\frac{\Lambda^2}{4}{(1-\mu)}^2 t^2 r^{2(1-\mu)}
\right) .
\label{c10}\end{eqnarray}
If we compare the exact non-perturbative limit (\ref{c8}) with what is given
by the first order approximation. i.e.
\beq |Z^{(0)}| = 1 \leftrightarrow r^{2(1-\mu)} \sim \frac{4}{\Lambda}
\label{c11}\eeq
we note that the correct value is almost obtained, and that the
following perturbative terms give only relatively small corrections.

To describe the scattering of point sources the monodromies must be
non-abelian between them, in such a way that their fixed points are
distinct for each particle:
\beq  Z \buildrel {1} \over \rightarrow
\frac{ a^1 Z + b^1 }{ \overline{b}^1 Z +
\overline{a}^1 } \ \ \ \ \ \ \ \ \ \ \ \
 Z \buildrel {2} \over \rightarrow
\frac{ a^2 Z + b^2 }{ \overline{b}^2 Z +
\overline{a}^2 } . \label{c12}\eeq
With the coefficients of the transformations defined as
\begin{eqnarray}
a^i & = & cos \pi \mu^i - i \gamma^i sin \pi \mu^i \nonumber \\
b^i & = & - i \gamma^i \overline{V}^i sin \pi \mu^i ,
\label{c13}\end{eqnarray}
the corresponding fixed points are defined by
\beq Z_i^{(0)} = - \frac{\gamma^i \overline{V}^i }{1+ \gamma^i} .
\label{c14}\eeq
Let us do the case of head-on collision,  choosing the coefficients
given in eq. (\ref{c13}) in the following way
\begin{eqnarray}
a_1 & = & cos \pi \mu_1 - i \gamma_1 sin \pi \mu_1 \ \ \ \ \ \ \ \ \
a_2 = cos \pi \mu_2 - i \gamma_2  sin \pi \mu_2 \nonumber \\
b_1 & = & - i \gamma_1 |v_1| sin \pi \mu_1 \ \ \ \ \ \ \ \ \
b_2 =  i \gamma_2 |v_2| sin \pi \mu_2 .
\label{c15}\end{eqnarray}
We can parameterize the coefficients only in terms of rapidities,
without extra phases, i.e.
\begin{eqnarray}
\gamma_1 & = & ch ( 2 \sqrt{\Lambda} \lambda_1 ) \ \ \ \ \ \ \ \ \
\gamma_2 = ch ( 2 \sqrt{\Lambda} \lambda_2 ) \nonumber \\
\gamma_1 |v_1| & = & sh ( 2 \sqrt{\Lambda} \lambda_1 ) \ \ \ \ \ \ \ \ \
\gamma_2 |v_2| = sh ( 2 \sqrt{\Lambda} \lambda_2 ) .
\label{c16}\end{eqnarray}
Developing the coefficients $a_k$ e $b_k$  in powers of the
perturbative parameter, the cosmological constant $\Lambda$,
\begin{eqnarray}
a^k & = & a^k_{(0)} + \Lambda  a^k_{(1)} + \Lambda^2  a^k_{(2)} + O(\Lambda^3)
\nonumber \\
b^k & = & \sqrt{\Lambda} ( b^k_{(0)} + \Lambda  b^k_{(1)} + \Lambda^2
b^k_{(2)} + O(\Lambda^3) ) \ \ \ \ \ \ k = 1,2
\label{c17}\end{eqnarray}
we find at the lowest orders in $\Lambda$
\begin{eqnarray}
a_1  & = & e^{-i \pi \mu_1} - 2 i \Lambda  sin \pi \mu_1
\lambda^2_1 + O(\Lambda^2) \ \ \ \ \ \ \
a_2 = e^{-i \pi \mu_2} - 2 i \Lambda  sin \pi \mu_2
\lambda^2_2 + O(\Lambda^2)
\nonumber \\
b_1 & = & - 2 i  \sqrt{\Lambda}  \left( \lambda_1 + \Lambda
\frac{2 \lambda^3_1}{3} + O(\Lambda^2) \right) sin \pi \mu_1  \ \ \ \ \ \
b_2 = + 2 i \sqrt{\Lambda}  \left( \lambda_2 + \Lambda
\frac{2 \lambda^3_2}{3} + O(\Lambda^2) \right) sin \pi \mu_2 . \nonumber \\
& & \label{c18}\end{eqnarray}

The system is defined non perturbatively by an invariant mass ${\cal
M}$ characterizing the system at great distances as a one-body metric.
Speaking of limit of great distances is a little misleading since the
metric in conformal gauge is valid only on a finite size region,
however in a perturbative expansion, the limit coming from re-summing
the perturbative series is not visible.

The invariant mass defined by
\beq cos \pi {\cal M} = cos \pi \mu_1 cos \pi \mu_2 -
ch (\sqrt{\Lambda} ( \lambda_1 + \lambda_2 ))  sin \pi \mu_1 sin \pi \mu_2 ,
\label{c19}\eeq
is a real number included between $0$ and $1$,
implying some non-perturbative limits on the
values of the constants of motion:
\beq ch ( \sqrt{\Lambda} ( \lambda_1 + \lambda_2 )) \leq
\frac{1+cos \pi\mu_1 cos \pi \mu_2 }{sin \pi \mu_1 sin \pi \mu_2 } .
\label{c19b} \eeq

   In the computation of the
perturbative fields it is useful to verify that the perturbative terms
can be recombined to give $\cal M$, confirming that the complete
solution, which we are not able to give explicitly, is deeply dependent
on it. In fact, as in the one-body case there is another singularity,
characterizing the $Z$-field at
infinity in the unphysical region, whose mass must be identified with
$\cal M$, as we shall see later on.

Before analyzing the two-body problem, we must notice that the
cosmological constant produces by itself a sort of background field,
which is obtained in the massless limit $\mu \rightarrow 0$:
\begin{eqnarray}
Z & = & \sqrt{1-a^2} \frac{\sqrt{\Lambda}}{4}  z \left[
1 + \frac{4 r^{-2}}{\Lambda} \left(
1- \sqrt{ 1 - \frac{\Lambda}{2} \left( \frac{ 1+a^2 }{ 1-a^2 } \right)
 r^2 + \frac{\Lambda^2}{16}  r^4 } \right) \right]
\nonumber \\
sin T & =& a \frac{1 + \frac{\Lambda}{4} r^2 }{
1 - \frac{\Lambda}{4} r^2 } \ \ \ \ \ \ \ \ a = sin (
\sqrt{\Lambda} t ) .
\label{c20}\end{eqnarray}
Therefore at each order the development of the background metric
gives rise to a contribution, which is divergent at infinity. This is
made smoother by the presence of the sources, that at infinity are
seen as a unique central body with total mass given by the invariant mass
${\cal M}$. Since we have to fix the boundary conditions we choose to
require that the analytic functions that naturally arise from the
integration are regular around the particles and have a behavior at
infinity that is no more divergent than a one-body metric with the mass
given by the invariant mass $\cal M$:
\begin{eqnarray}
T & \buildrel {r \rightarrow  \infty} \over \longrightarrow &
\sqrt{\Lambda} (1-{\cal M}) t \left( 1 + \frac{\Lambda}{2}
 r^{2(1-{\cal M})}   + O({\Lambda}^2) \right) \nonumber \\
Z & \buildrel {r \rightarrow  \infty} \over \longrightarrow
 & \frac{\sqrt{\Lambda}}{2} \frac{z^{1-{\cal M}}}{\sqrt{1-\widetilde{a}^2
(t)}} \left( 1 + \frac{\Lambda^2}{4}{(1-{\cal M})}^2 t^2 r^{2(1-{\cal M})}
\right) ,
\label{c21}\end{eqnarray}
where $\widetilde{a} (t) = sin [ (1-{\cal M}) t ]$. It can be necessary to
adjust the solution asymptotically with a boost of  $Z$,
( $a_{\infty} = ch(2\sqrt{\Lambda} \lambda_{\infty} )$ ,
$b_{\infty} = sh(2\sqrt{\Lambda} \lambda_{\infty}) $ ), implying in
practice a redefinition of $Z$ with respect to the static case (\ref{c21}):
\beq Z \rightarrow \frac{a_{\infty} Z + b_{\infty}}{b_{\infty} Z +
a_{\infty}} .
\label{c22}\eeq

In (\ref{c15}) we have supposed that at the lowest order
the cut is purely a rotation,
\begin{eqnarray}
a_{(0)}^k & = & e^{-i \pi \mu_k} \nonumber \\
b_{(0)}^k & = & 2 i (-)^k \lambda_k sin \pi \mu_k
\ \ \ \ k = 1,2 ,
\label{c23}\end{eqnarray}
then the monodromy conditions reduce at the same order to
\beq Z^{(0)} \buildrel {k} \over \rightarrow
e^{-2i \pi \mu_k } Z^{(0)} + (-)^k 4 i \lambda_k  e^{-i \pi \mu_k}
sin \pi \mu_k , \label{c24}\eeq
from which we recover the static solution defined in the rescaled variable
$\zeta = ( z - \xi_2 ) / ( \xi_1 - \xi_2 )$ by
\beq Z^{(0)} = k_2 \int^\zeta_0 d\zeta \zeta^{-\mu_1}
{(1-\zeta)}^{-\mu_2}
+ k_1 . \label{c25}\eeq

Eq. (\ref{c24}) determines $k_1 $ and $k_2$
\begin{eqnarray}
k_1 & = & - 2 \lambda_1  \nonumber \\
k_2 & = & (1-\mu_1-\mu_2 ) \xi_{12}^{1-\mu_1-\mu_2}
=  \frac{2( \lambda_1 + \lambda_2 )}{B(1-\mu_1,1-\mu_2)}
+ O(\Lambda) .
\label{c26}\end{eqnarray}
The fact that it is possible to relate the physical distance
$\xi_{12}$ in terms of the bosonic distance $Z^{(0)} (1) - Z^{(0)}
(0)$, is connected with the requirement that the mapping $Z^{(0)}$
reduces to the identity mapping $z$ in the massless limit.

In practice, from the one body case, we can deduce what is the
approximate figure of the patch on which the solution is valid,
defined by the equation

\beq {|Z^{(0)}|}^2 = \frac{4}{\Lambda} . \label{c27}\eeq
If we insist too much  on the validity of eq. (\ref{c27}),
we find some perturbative limit on the constant of motion
$\lambda_i \leq\frac{2}{\sqrt{\Lambda}}$, which really doesn't exist,
and therefore we deduce that this approximate equation is valid only
for small values of $\lambda_i$. However more subtle limits on the constants
of motion appear if we require that the asymptotical behavior is
related to a particle, and not to a tachyon, as shown in eq. (\ref{c19b}).

Let us rewrite the zero order solution in terms of an
hypergeometric function:
\begin{eqnarray}
 Z^{(0)} & = & k_2 \int^{\zeta}_0 d \zeta \zeta^{-\mu_1} {(1-\zeta)}^{-\mu_2}
+ k_1 \nonumber \\
& = & k_2 \frac{\zeta^{1-\mu_1}}{1-\mu_1}
F(\mu_2, 1-\mu_1, 2-\mu_1 ; \zeta ) + k_1
\buildrel {\zeta \rightarrow \infty} \over \longrightarrow
z^{1-\mu_1-\mu_2} .
\label{c28}\end{eqnarray}
The analytic part of the solution for $Z$ can be directly generalized
at all orders representing it as a ratio of hyper-geometric functions
\beq \widetilde{Z} = coth(\sqrt{\Lambda} ( \lambda_1+\lambda_2 ))
 \ \zeta^{1-\mu_1} \frac{ \widetilde{F} \left(\displaystyle{ \frac{{\cal
M}-\mu_1+\mu_2}{2}, 1 + \frac{-{\cal M}-\mu_1+\mu_2}{2}}, 2-\mu_1;
\zeta \right) }{\widetilde{F} \left( - 1 + \displaystyle{ \frac{{\cal
M}+\mu_1+\mu_2}{2}, \frac{-{\cal M}+\mu_1+\mu_2}{2} } , \mu_1;
\zeta \right) } , \label{c29}\eeq
where the symbol $\widetilde{F}$ denotes a novel normalization of the
hyper-geometric function ( see also \cite{a6} ):
\beq \widetilde{F} (a,b,c;z) = \frac{\Gamma(a) \Gamma(b)}{\Gamma(c)}
F (a,b,c;z) . \label{c30}\eeq
More generally, the solution will be defined by a Lorentz
transformation in order to produce the right fixed points of the $Z$-monodromy
\beq Z= \frac{\widetilde{Z} + Z(0)}{ 1+ \overline{Z(0)} \widetilde{Z} } ,
\label{c31}\eeq
which, in the special case of head-on collision, become
\beq Z(0) = - \frac{\gamma_1 V_1}{1+\gamma_1} = - th ( \sqrt{\Lambda}
\lambda_1 ) \ \ \ \
Z(1) = \frac{\gamma_2 V_2}{1+\gamma_2} = th ( \sqrt{\Lambda}
\lambda_2 ) \ \ \ \
\widetilde{Z}(1) = th(\sqrt{\Lambda} (\lambda_1+\lambda_2)) .
\label{c32}\eeq

It is easy to verify that the first order solution ( eq. (\ref{c28}) )
is obtained with the following limit on the invariant mass
\beq \Lambda \rightarrow 0  \equiv {\cal M} \rightarrow \mu_1 + \mu_2
+ 2 \Lambda {( \lambda_1 +
\lambda_2 )}^2 \frac{ sin \pi \mu_1 sin \pi \mu_2 }{ \pi sin \pi ( \mu_1 +
\mu_2 )} + O (\Lambda^2 ) \label{c33}\eeq
once that the particular values of head-on collision are substituted.
From the definition of the hyper-geometric function it is clear that
there is another singularity at infinity, specular to the ones
situated in $\zeta = 0$ and $\zeta = 1$, whose cut is related to the
invariant mass ${\cal M}$.

The scale $\xi_{12}$ appear to be undetermined in the exact analytic
solution. As in the lowest order in $\Lambda$, corresponding to pure
gravity, there is no way to determine it unless we require that the
solution respects a given asymptotic behavior, as we have discussed
before deriving eqs. (\ref{b21}) and (\ref{b22}).

To give the contribution to the geodesic equations, it is useful to
make an asymptotic development of the $Z$ solution in the limit $\zeta
\rightarrow \infty$:
\beq \widetilde{Z} \buildrel {\zeta \rightarrow \infty} \over
\longrightarrow \frac{a_1 + a_2 \zeta^{1-{\cal M}} }{a_3 + a_4
\zeta^{1-{\cal M}} } , \label{c34}\eeq
with the coefficients defined as:
\begin{eqnarray}
a_1 & = & coth( \sqrt{\Lambda} ( \lambda_1 + \lambda_2 )) \frac{
\Gamma(b) \Gamma ( a-b ) }{ \Gamma ( c-b )} \ \ \ \ \ \
a_2 =  coth( \sqrt{\Lambda} ( \lambda_1 + \lambda_2 )) \frac{
\Gamma(a) \Gamma ( b-a ) }{ \Gamma ( c-a )} \nonumber \\
a_3 & = & \frac{\Gamma(b-c+1) \Gamma ( a-b ) }{ \Gamma ( 1-b )} \ \ \ \ \ \
a_4 = \frac{\Gamma(a-c+1) \Gamma ( b-a ) }{ \Gamma ( 1-a )} \nonumber \\
a & = & \frac{ {\cal M} - \mu_1 + \mu_2}{2} \ \ \ \ \
b = 1 + \frac{ - {\cal M} - \mu_1 + \mu_2}{2} \ \ \ \ \
c = 2 - \mu_1 .
\label{c35}\end{eqnarray}
This development can be put in the form of a Lorentz transformation
with respect to the static case
\beq \widetilde{Z} = \frac{ a_{\infty} Z + b_{\infty} }{b_{\infty} Z
+ a_{\infty} }
\ \ \ \ \ \ \ \ Z \buildrel{\zeta \rightarrow \infty} \over \longrightarrow
\frac{\sqrt{\Lambda}}{2} \frac{z^{1-{\cal M}}}{\sqrt{ 1- \widetilde{a}^2(t)}} ,
\label{c36}\eeq
where we have taken into account the fact that the asymptotic metric
contains terms dependent explicitly on time ( $\widetilde{a}
(t) = sin ( \sqrt{\Lambda} ( 1 - {\cal M} ) t ) $ ). The solution for
the scale $\xi_{12}$ becomes then:
\beq \xi_{12}^{1-{\cal M}} = \frac{2\sqrt{ 1- \widetilde{a}^2(t)}}
{\sqrt{\Lambda}}  \frac{a_2}{a_3} \label{c37}\eeq
as the following identity  $a_1 a_2 = a_3 a_4 $ is valid.

At the next order $\Lambda^2$ we find that the field time dependence
comes directly from the equations of motion and not only from the
boundary conditions. To compute its contribution we must first know
$T^{(1)}$.

From the asymptotic behavior ( eqs. (\ref{c21}) and (\ref{c22}) ) we
deduce that:
\beq T^{(0)} = ( 1- {\cal M} t ). \label{c38}\eeq
Instead, the equation of motion for $T^{(1)}$ can be integrated giving
\beq \partial_z T^{(1)} = \frac{T^{(0)}}{2} \partial_z Z^{(0)} (
\overline{Z}^{(0)} + h(\zeta) ). \label{c39}\eeq
In the limit $\mu_i \rightarrow 0$ we impose that the contribution of
the extra function $h(\zeta)$ vanishes, so to recover the background field
(\ref{c20}). Let us define $h(\zeta)$ in such a way that $\partial_z T^{(1)}$
is a meromorphic function, which therefore requires that the combination
$\overline{Z}^{(0)} + h(\zeta)$ transforms as $\partial_{\overline{z}}
\overline{Z}^{(0)}$ under all monodromies, from which we deduce the
monodromy properties of $h(\zeta)$ as follows:
\beq h(\zeta) \buildrel {k} \over \rightarrow
e^{2i\pi\mu_k} h(\zeta) + 4i (-)^k \lambda_k e^{i\pi\mu_k} sin \pi \mu_k
\label{c40}\eeq

Adding the condition that $h(\zeta) \rightarrow 0$ in the limit of
small masses, $\mu_i \rightarrow 0$, we obtain the solution,
perturbative in the mass parameters,
\beq h(\zeta) \simeq - 2\lambda_1 \mu_1 log \zeta + 2\lambda_2 \mu_2 log ( 1 -
\zeta) . \label{c41}\eeq
This behavior, divergent around the particles, is still acceptable,
because the logarithmic divergence is cancelled by $z$-integration
and $T^{(1)}$ is again well defined around the particles.

Let us also notice that in the case of one particle $h(\zeta)$ is
constrained by
\beq h(\zeta) \buildrel {\mu_2 \rightarrow 0} \over \longrightarrow
2 \lambda_1 ( 1 - C z^{\mu_1} ) . \label{c42}\eeq
hence the solution for $T^{(1)}$ must be for one particle with
rapidity $\lambda_1 \neq 0$,
\beq T^{(1)} = \frac{T^{(0)}}{2} \left[ ( Z^{(0)} + 2 \lambda_1 ) (
\overline{Z}^{(0)} + 2 \lambda_1 ) - 2\lambda_1  C ( 1- \mu_1 ) ( z +
\overline{z} ) \right] + T^{(1)} (t) .\label{c43}\eeq

We can generalize eq. (\ref{c41}) to the case of any masses
choosing $h(\zeta)$ as
\beq h(\zeta) = A_1 \int^\zeta_1 d\zeta \zeta^{\mu_1-1} {(1-\zeta)}^{\mu_2} +
 A_2 \int^\zeta_0 d\zeta \zeta^{\mu_1} {(1-\zeta)}^{\mu_2-1} . \label{c44}\eeq
The monodromy conditions (\ref{c40}) are satisfied by the following positions:
\beq A_1 = - \frac{2 \lambda_1}{B(\mu_1,1+\mu_2)}
\ \ \ \ \ \ \ A_2 = - \frac{2 \lambda_2}{B(1+\mu_1,\mu_2)} , \label{c45}\eeq
that in the small mass limit reproduces the solution (\ref{c41}).

Therefore we can define the integrated field $T^{(1)}$ as
\begin{eqnarray}
 T^{(1)} & = & \frac{T^{(0)}}{2} \left[ Z^{(0)} \overline{Z}^{(0)} +
\int^z_{\xi_1} dz \partial_z Z^{(0)} h(\zeta) +
\int^{\overline{z}}_{\overline{\xi}_1} d{\overline{z}} \partial_{\overline{z}}
\overline{Z}^{(0)} \overline{h}(\overline{\zeta}) \right] + T^{(1)}(t)
\nonumber \\
& = & \frac{T^{(0)}}{2} \left[ ( Z^{(0)} - 2 {(-)}^k \lambda_k )
( \overline{Z}^{(0)} - 2 {(-)}^k \lambda_k ) + \int^z_{\xi_1} dz
\partial_z Z^{(0)} ( h(\zeta) + 2 {(-)}^k \lambda_k ) + \right. \nonumber \\
& + & \left. \int^{\overline{z}}_{\overline{\xi}_1} d{\overline{z}}
\partial_{\overline{z}} \overline{Z}^{(0)}
( \overline{h}(\overline{\zeta}) + 2 {(-)}^k
 \lambda_k ) \right] + T^{(1)}_k (t)
\ \ \ \ \ \ \ \ k = 1,2 . \label{c46}\end{eqnarray}
We have added the second line to make explicit the property that
$T^{(1)}$ is automatically monodromic around both particles, without
need to introduce logarithmic terms to adjust an eventual translation
monodromy, which fortunately is absent.

Starting from the knowledge of $T^{(1)}$ we can deduce, using the
first integral (\ref{b18}) constraints for the solution of the $Z^{(2)}$ field,
\beq \partial_{\overline{z}} Z^{(2)} = \frac{ {(
\partial_{\overline{z}}
T^{(1)} )}^2 }{ \partial_{\overline{z}} \overline{Z}^{(0)} } =
\frac{ {( T^{(0)} )}^2 }{4} { ( Z^{(0)} + \overline{h}(\zeta) ) }^2
\partial_{\overline{z}} \overline{Z}^{(0)} . \label{c47}\eeq

This first integral is automatically solution of the perturbative
expansion of eq. (\ref{b17}), which remains also non linear in the unknown
$\partial_{\overline{z}} Z^{(2)}$ after the development in $\Lambda$:
\beq \partial_z \partial_{\overline{z}} Z^{(2)} = T^{(0)} \partial_z
Z^{(0)} \sqrt{ \partial_{\overline{z}} \overline{Z}^{(0)}
\partial_{\overline{z}} Z^{(2)} } . \label{c48}\eeq

The other first integral, which is relative to $N(z)$, gives at the
lowest order in $\Lambda$:
\beq N(z) = \Lambda \frac{ T^{(0)} }{2} k_2 \left( \frac{A_1}{\zeta} +
\frac{A_2}{1-\zeta} \right) \label{c49}\eeq
and we check that is a meromorphic function with simple poles.

The solution for $Z^{(2)}$ can be decomposed in an analytic part and a
non-analytic one, satisfying eq. (\ref{c47}):
\beq Z^{(2)} = Z^{(2)}_a (z) +  Z^{(2)}_n (z, \overline{z}) \label{c50}\eeq
While the analytic part  $ Z^{(2)}_a (z)$ is determined by the
development of the general solution (\ref{c29}), the non-analytic
part must satisfy only the homogeneous part of the monodromies
\beq  Z^{(2)}_n (z, \overline{z}) \buildrel {k} \over \longrightarrow
e^{-2i \pi \mu_k }  Z^{(2)}_n (z, \overline{z}) . \label{c51}\eeq

When we integrate $\partial_{\overline z} Z^{(2)}$ we have at
disposition an arbitrary polydromic function $f(z)$, to be fixed in
order to make $Z^{(2)}_n (z, \overline{z})$ again covariant under the
rule (\ref{c51}).

A possible solution is
\begin{eqnarray}
Z^{(2)}_n (z, \overline{z} ) & = & \frac{T^{(0) 2}}{4} \left[ { (
Z^{(0)} + 2 \lambda_1 ) }^2 ( \overline{Z}^{(0)} + 2 \lambda_1 ) + \right.
 2 ( Z^{(0)} + 2 \lambda_1 ) \int^{\overline{z}}_{\overline{\xi}_1}
d \overline{z}
\partial_{\overline{z}} \overline{Z}^{(0)} ( \overline{h} (
\overline{\zeta} ) - 2 \lambda_1 ) + \nonumber \\
& + & \left. \int^{\overline{z}}_{\overline{\xi}_1} d \overline{z}
\partial_{\overline{z}} \overline{Z}^{(0)} {( \overline{h} (
\overline{\zeta} ) - 2 \lambda_1 )}^2 + f(\zeta) \right] = \nonumber \\
 & = & \frac{T^{(0) 2}}{4} \left[ { (
Z^{(0)} - 2 \lambda_2 ) }^2 ( \overline{Z}^{(0)} - 2 \lambda_2 ) + \right.
 2 ( Z^{(0)} - 2 \lambda_2 ) \int^{\overline{z}}_{\overline{\xi}_2}
d \overline{z}
\partial_{\overline{z}} \overline{Z}^{(0)} ( \overline{h} (
\overline{\zeta} ) + 2 \lambda_2 ) + \nonumber \\
& + & \left. \int^{\overline{z}}_{\overline{\xi}_2} d \overline{z}
\partial_{\overline{z}} \overline{Z}^{(0)} {( \overline{h} (
\overline{\zeta} ) + 2 \lambda_2 )}^2 + g(\zeta) \right] ,
\label{c52}\end{eqnarray}
in which the first expression is automatically covariant under the
first particle if
\beq f(\zeta) \buildrel {1} \over \longrightarrow e^{-2\pi\mu_1}
f(\zeta)  ,
\label{c53}\eeq
while the second expression is covariant around the second particle if
\beq g(\zeta) \buildrel {2} \over \longrightarrow e^{-2\pi\mu_2}
g(\zeta). \label{c54}\eeq
Developing these two formulas we find that
\begin{eqnarray} f(\zeta) & = &
g(\zeta) - 2 ( \lambda_1 + \lambda_2 ) { ( Z^{(0)} )}^2 + 2 C_1
Z^{(0)} + C_2 \nonumber \\
& = & g(\zeta) - 2 ( \lambda_1 + \lambda_2 ) { ( Z^{(0)} - 2 \lambda_2
)}^2 + 2 \widetilde{C}_1 ( Z^{(0)} - 2 \lambda_2 ) + \widetilde{C}_2
\nonumber \\
\widetilde{C}_2 & = & C_2 + 4 \lambda_2 C_1 + 8 \lambda^2_2 ( \lambda_1 +
\lambda_2 ) ,
\label{c55}\end{eqnarray}
where to know $C_1$ and $C_2$ we must solve the following integrals
\beq C_1 =  \int^{\xi_2}_{\xi_1} dz \partial_z Z^{(0)} h(\zeta) \ \ \ \ \
C_2 =  \int^{\xi_2}_{\xi_1} dz \partial_z Z^{(0)} h^2(\zeta) . \label{c56}\eeq

The simplest solution is in fact
\begin{eqnarray}
f_0(z) & = & \Delta_1 \int^z_{\xi_1} dw \partial_w Z^{(0)} (w) \int^{w}_{\xi_1}
dv \partial_v Z^{(0)}(v) \int^{\zeta = (v-\xi_2)/\xi_{12}}_0
d\zeta \zeta^{\mu_1} {(1-\zeta)}^{\mu_2 -1} +
\Delta_2 ( Z^{(0)} + \lambda_1 ) \nonumber \\
\Delta_1 & = & - \frac{ 2 (\lambda_1 + \lambda_2)}{B(1+\mu_1,\mu_2)}
\nonumber \\ \Delta_2 & = &
\frac{\widetilde{C}_2}{(\lambda_1+\lambda_2)} + \frac{2}{B(1+\mu_1,\mu_2)}
\int^{\xi_2}_{\xi_1} dw \partial_w Z^{(0)} (w) \int^{w}_{\xi_1}
dv \partial_v Z^{(0)}(v) \int^{\zeta = (v-\xi_2)/\xi_{12}}_0
d\zeta \zeta^{\mu_1} {(1-\zeta)}^{\mu_2 -1} . \nonumber \\
& & \label{c60}\end{eqnarray}
To match the background metric (\ref{c20}) we can always add to the
particular solution $f_0(z)$ terms of the type:
\beq f(z) = f_0(z) + ( A + B w + C w^2 ) \zeta^{\mu_1}
{(1-\zeta)}^{\mu_2} .
\label{c59}\eeq

At the level of geodesic equations we notice that the condition
(\ref{c32}) is solved for an arbitrary scale $\xi_{12}$ and
that it doesn't give rise
to new constraints, while the criterium of reproducing a certain
asymptotic behavior is already satisfied by the first order of
eq. (\ref{c37}), i.e. by eq. (\ref{c26}). If we continue
perturbation theory it is possible that other constraints result from
the requirement that the residue of the simple poles of $N(z)$ has the
following form, valid at all orders:
\beq N(z) \sim \Lambda {(\xi_{12})}^{1-{\cal M}} \left( \frac{A_1
(t)}{\zeta} +
\frac{A_2 (t) }{1-\zeta} \right) \label{c63}\eeq
with the invariant mass ${\cal M}$ that replaces the sum of masses.

We cannot forget that we are treating a particular case of scattering
at zero angular momentum, in which the particles have to collide after
a finite time. This case is typically ill-defined from the
distributional point of view, as in the solution products of
distributions are expected to appear at a certain time, while the
geodesic limit is still well defined. A verification of the
consistency of this solution can be made with a
perturbative computation at non-vanishing angular momentum, in which
case it is no more possible to reduce the general $O(2,2)$ monodromies
to $O(2,1)$.

\section{Conclusions}

We have deduced the general equations for the immersion $X^A$ that
governs the scattering of point sources coupled to $AdS$-gravity. We
have found the complete solution for the analytic part of the fields
and a partial one for the non-analytic part. The choice of conformal
gauge allows to study the scattering problem with instantaneous
propagation of the fields avoiding the difficulties connected with the retarded
potentials. This gauge is globally defined for the scattering of particles,
but it is not for what concerns the scattering of black holes, where
its validity is reduced to the region external to the horizons.

The scattering of particles is governed by the composition of
monodromies, which gives rise to two invariant masses \cite{a9}
; someone may object that
these are not relevant for the solution of the fields, since in conformal
gauge there is a physical limit on the values of spatial coordinates
and it is not possible to see the particles as a unique body at
great distance from them. However such a limit is non-perturbative with
respect to the
cosmological constant and at a perturbative level the fields do have
infinite extension, and the leading contribution at great distances of
each perturbative order must be dominated by these two invariant masses,
which in our particular case of head-on collision coincide. The
physical scale is produced only  re-summing the perturbative series.
At a non-perturbative level, the two invariant masses are still present,
since in the non-physical region of the $Z$ field it appears a specular image
of the $N$-body system with an unique singularity, defined by the two
invariant masses. Therefore, at least mathematically, this extra
singularity is important to give the parameterization of the solution.

It would be interesting to continue our study to include the
scattering of $BTZ$ black holes , where the equations (\ref{b12})
still remain valid outside their horizons. A similar problem has
already an analogy with the case of the scattering of spinning
particles in $(2+1)$-gravity \cite{a8}, in which the problem of closed
time-like curves produces, in conformal gauge, some $CTC$ horizons
around the particles. Work is in
progress in this direction.

{\bf Acknowledgements}

I would like to thank A. Cappelli and M. Ciafaloni for useful
discussions.

\end{document}